\documentstyle[aps,prl,epsfig]{revtex}
\begin{document}
\draft
\title{Behavior of a polymer chain in a 
critical binary solvent
}
\author{${\mbox{Michael\ Stapper} }^{}$ and
${\mbox{Thomas A.\ Vilgis}}^{}$}
\address{
Max-Planck-Institut f\"ur Polymerforschung,
Postfach 3148, D-55021 Mainz, Germany
}
\date{\today}
\maketitle

\begin{abstract}
We present a field-theoretic renormalization group analysis of a
polymer chain immersed in a binary good solvent close to its critical
demixing point. We first show that this problem can be mapped on a
bicritical field theory, i.e. a $(\Phi^{2})^{2}$-model with a mass anisotropy.
This implies that the end-to-end distance of the polymer is now controlled
by a new critical exponent $\nu_{B}$ related to the quadratic mass anisotropy
operator $B$. To show this we solve the RG equation and calculate explicitly 
the exponents and the mean end-to-end length of the chain.
\end{abstract}

In this paper we consider the critical behavior of a polymer chain in a 
critical 
fluid~\cite{1-4}. This problem is quite poorly understood, because its a
coupled system with two macroscopic length scales. The physics is determined 
by an interplay of the chain length and the correlation length of the
critical fluid, which may undergo a phase separation at a certain temperature.
The polymer in solution and the critical fluid are separately successfully
described by scaling and renormalization group theories. The separation of 
microscopic and macroscopic length scales leads there to universality. 
That means
the main properties of the system are independent of microscopic details,
which only change prefactors in the physical quantities and make analysis
of experimental results, at least in their limiting behavior straightforward.
We will show that also for our coupled system the application of the 
RG is possible and yields a description of the conformation of 
the chain. 
First insight beyond mean field theory was brought in by Brochard and
de Gennes~\cite{Br&G} by the use of scaling theory and by the simulation of
Magda et al~\cite{Ma}, but still there is a lack
of a quantitative 
investigation of the chain conformation near the consolute point of 
the solvent. The start point of such an investigation is the
settled up field theory by Vilgis et al~\cite{Vilgis}.

The polymer chain of configuration ${\bf R}(s)$ is modeled by an
Edwards-Wiener functional~\cite{Doi}
\begin{equation}
{\cal H}_{0}=\frac{d}{2l^{2}}\int_{0}^{L}ds \left( \frac{\partial {\bf R}(s )}
{\partial s}\right)^{2} +
 v \int_{0}^{L}\int_{0}^{L}ds ds'
\delta ({\bf R}(s)-{\bf R}(s'))\quad ,
\end{equation}
where $s$ is the dimensionless contour variable, $l$ the 
Kuhn segment length, $d$ the space dimension, $v$ the parameter 
of the bare short ranged excluded volume interaction, $N$ is the number of
segments and $L=l N$ the contour length of the chain. 
The reduced Hamiltonian of the fluid can be given in terms of 
three short range pair-potentials between the 
good solvent molecules
\begin{equation}
 H_{f}=\sum_{\sigma , \tau}\sum_{i,j} V^{\sigma\tau} 
({\bf r}_{i}^{\sigma}-{\bf r}_{j}^{\tau}) 
\end{equation}
where $\sigma , \tau =1,2$ and $i,j$ label the molecules. The interaction 
between the chain and the fluid is described as 
\begin{equation}
{\cal H}_{i}=
\int_{0}^{L}ds \sum_{\sigma =1,2}\sum_{j} v_{m\sigma}
\delta ({\bf R}(s)-{\bf r}_{j}^{\sigma})\quad ,
\end{equation}
where $v_{m\sigma}$ characterizes the short range excluded volume interaction
between the monomers and the different species of the fluid. If their solvent
quality is slightly different, that means $v_{m1} >  v_{m2     }$
or $v_{m1}<v_{m2}$, there is
preferential affinity, the chain is likely to be surrounded by solvent $2 $
or solvent $1$. After the use of collective density fields~\cite{Vilgis}
 for the fluid
and the chain we get the equivalent complete field theoretical 
Hamiltonian
\begin{eqnarray}
{\cal H} &=& \int d^{d}r\left\{\frac{1}{2}(\nabla {\bf \phi}(r))^{2} 
+\frac{1}{2}y_{0} {\bf \phi }^{2}(r) +
\frac{1}{2}(\nabla { \psi}(r))^{2}+\frac{1}{2}\tau_{0}\psi^{2}(r)+\right.
\nonumber\\
&&\left.+\frac{w_{0}}{2}{\bf \phi}^{2}(r)\psi
+\frac{g_{1}}{4!}({\bf \phi}^{2}(r))^{2}+
\frac{g_{2}}{4!}\psi^{4}(r)
+2\frac{g_{3}}{4!}{\bf \phi}^{2}(r)\psi^{2}(r)\right.\quad .
\end{eqnarray}
The polymer field ${\bf \phi }$ is an $\tilde M$ component field with 
$\tilde M\to 0$
after the evaluation of the perturbation expansion. 
The quantity ${\bf \phi }^{2}$ is related to the monomer
density. This follows from the well known de Gennes transformation. Whereas
$\psi$ is a scalar field which describes the relative concentration 
fluctuations of one of the fluid constituents.
The term ${\bf \phi}^{2}\psi^{2}$ originates from the incompressibility 
condition~\cite{Vilgis} and is important for the physics: In the case of
preferential affinity ($w_{0}> 0$) it 
re-swells the chain again in the critical region
of the consolute point~\cite{Br&G,Ma}. 
Whereas in the case of no preferential affinity
it is the lowest order coupling between fluid and polymer. It is interesting to
note that the same Hamiltonian but with an imaginary coupling $w_{0}$
appears in the modeling of a single screened polyelectrolyte chain, i.e. a 
Debye H\"uckel
chain with corrections~\cite{LS}.

In the following we consider the case with no preferential affinity, i.e. 
the cubic coupling $w_{0}=0$. We may write the Hamiltonian as one with a
composite $M=\tilde M +1$ 
component field, with $M\to 1$ due the polymer limes. The
fixed  point of such a system is for $M<4$ the Heisenberg fixed point with
all corresponding renormalized 
quartic couplings equal $u_{1}^{*}=u_{2}^{*}=u_{3}^{*}=
u^{*}(M)$~\cite{Am}. Due to universality we are allowed to set these
couplings already equal in the original (renormalized) Hamiltonian~\cite{Am}
and we get
\begin{eqnarray}
\label{b1}
{\cal H} &=& \int d^{d}r\left\{
\frac{1}{2}(\nabla {\bf \Phi}(r))^{2} +\frac{1}{2}y_{0}{\bf \phi}^{2}(r) +
\frac{1}{2}\tau_{0}{\psi}^{2}(r)+\frac{u_{0}}{4!}
({\bf \Phi}^{2}(r))^{2}\right\}
\quad ,
\end{eqnarray}
with ${\bf \Phi}=({\phi},\psi)$. 
This is exactly the Hamiltonian of a bicritical field
theory and well 
investigated by Amit and Goldschmidt~\cite{Am&G}, but now with $y_{0}$
as a monomer chemical potential, 
an inverse Laplace variable. To analyze the general 
behavior of the end to end distance, we use the so called soft expansion.
That means we write~(\ref{b1}) in the form~\cite{Am&G},
\begin{equation}
\label{b2}
{\cal H} = \int d^{d}r\left\{
\frac{1}{2}(\nabla {\bf \Phi}(r))^{2} +\frac{1}{2}t_{0} {\bf \Phi}^{2}(r) 
+\frac{u_{0}}{4!}({\bf \Phi}^{2}(r))^{2}-\frac{g_{0}}{2}B(r)\right\}
\quad ,
\end{equation}
with the definitions
\begin{eqnarray}
\label{b3o}
B(r) &=& 1/M \left((M-1)\psi^{2}(r)-{ \phi}^{2}(r)\right)\nonumber\\
t_{0} &=& 1/M (\tau_{0} + (M-1)y_{0}) \nonumber\\
g_{0} &=& \tau_{0} - y_{0}\quad .
\end{eqnarray}
The advantage of this splitting is 
that ${\bf \Phi}^{2}$ and $B$ are orthogonal under renormalization,
that means divergences which are generated by ${\bf \Phi}^{2}$ ($B$) 
insertions in vertex functions are absorbed completely by the corresponding
counterterms of ${\bf \Phi}^{2}$ ($B$) and these counterterms do not mix
with each other under renormalization. We
calculate the new exponent $\nu_{B}$ 
which is given by the anomalous dimension of the operator $B$
corresponding due to the limit $M\to 1$
to the monomer density. The other exponents $\eta$, $\nu$
and the $\beta$-function are the usual Heisenberg ones, which is obvious in
dimensional regularization~\cite{Am} (the masses $g_{0}$ and $t_{0}$ are then
allowed to set to zero). The upper 
critical dimension $d_{c}=4$ follows from a dimensional analysis of~(\ref{b2})
as for the symmetric Heisenberg model with $g_{0}=0$. . 
We show that
\begin{eqnarray}
\label{b3}
\nu_{B}&=& 1/(2-\gamma_{B})\nonumber\\
&=&(1+\epsilon/9+O(\epsilon^{2}))/2 \quad ,
\end{eqnarray}
and the cross over exponent is
\begin{eqnarray}
\label{b4}
\phi &=&\nu/\nu_{B}\nonumber\\
&=& 1+\epsilon/18+O(\epsilon^{2}) \quad ,
\end{eqnarray}
and 
$\nu=(1+\epsilon/6+O(\epsilon^{2})/2$, consistent with~\cite{Am&G}. 
But the soft
expansion~(\ref{b2}) is only useful for general considerations like the 
solution of RG-equations. Whereas for perturbation expansions of polymer
quantities it is not possible to use~(\ref{b2}), because an expansion in 
$g_{0}$ is
an expansion in an inverse Laplacian and therefore all back transformed
contributions are order by order zero.

Now we perform a general renormalization group
 analysis of our model given by equation~(\ref{b2}). Our
goal is to calculate the end to end distance of the polymer
\begin{eqnarray}
\label{b5}
<R^{2}>&=&-T_{{\cal L}}^{-1}\left(\partial_{q^{2}}
\tilde G^{(2)}(q)|_{q=0})\right)\left/
T_{{\cal L}}^{-1}\left(\tilde G^{(2)}(q=0)\right)\right. \nonumber\\
 &=& D(u) F(\tau,{\cal L})\quad ,
\end{eqnarray}
here is 
$G^{(2)}(r)=\int d^{d}q/(2\pi)^{d}\mbox{e}^{i{\bf q\cdot r}}\tilde G^{(2)}(q)=
{\cal N}\int{\cal D}{\bf \Phi}{\cal D}\psi \;
\Phi (r)\Phi (0)\mbox{e}^{-H({\bf \Phi},\psi)}$,
the propagator of the polymer, ${\cal N}$ is a normalization constant,
$T_{{\cal L}}^{-1}$ is the inverse Laplacian with respect to (wrt) 
${\cal L}=2lL=2l^{2}N/d$,
$\tilde G^{(2)}(q)$ 
is the Fourier transform of the renormalized chain propagator and
$D(u)$ is a non universal constant.

Due to the particular choice of $B$ we have a separate (diagonal)
renormalization scheme~\cite{Am&G}, 
which expresses the bare parameters in terms
of their renormalized ones:
\begin{eqnarray}
\label{b6}
t_{0}= Z_{\Phi^{2}}t, \;
g_{0}= Z_{B}g, \;
u_{0}=Z_{u}uS_{d}\mu^{\epsilon}
\end{eqnarray}
Here $\mu^{-1}$ is an external length scale and $S_{d}=\Omega_{d}/(2\pi)^{d}$
with $\Omega_{d}$ the volume of the unit sphere. 

We denote an $N$ point vertex function as $\Gamma^{(N)}$ and a composite one
with $N$ points and $L$, ${\bf \Phi}^{2}$  and $K$, $B$ insertions as
$\Gamma^{(N,L,K)}$. The wave-function renormalization factor is calculated
by expressing the renormalized vertex functions in terms of the bare ones as
\begin{equation}
\label{b7}
\Gamma_{R}^{(N)}=Z_{\Phi}^{N/2}\Gamma^{(N)} \quad .
\end{equation}
Whereas the renormalization of the composite ones give similar~\cite{Am&G}
\begin{equation}
\label{b8}
\Gamma_{R}^{(N,L,K)}=
Z_{\Phi}^{N/2}Z_{\Phi^{2}}^{L}Z_{B}^{K}\Gamma^{(N,L,K)} \quad .
\end{equation}
The renormalization group equations follow from an variation wrt the external
scale $\mu$ and are given by 
\begin{equation}
\label{b9}
\left[\mu\partial_{\mu}+
\beta\partial_{u}+\gamma_{\Phi^{2}}\tau\partial_{\tau}+\gamma_{B}g\partial_{g}
-\gamma_{\Phi}N/2\right]\Gamma^{(N)}_{R}=0\quad ,
\end{equation}
with the Wilson-functions $\beta=\mu{du}/{d\mu}$, $\gamma_{B/\Phi^{2}}
=-\beta 
({\partial\ln{Z_{B/\Phi^{2}}}}/{\partial u})$ 
and $\gamma_{\Phi}=\beta(\partial\ln{Z_{\Phi}}/\partial u)$.
The equivalent form in polymer variables is obtained by an inverse
Laplace Transformation with partial integration wrt $g$. The vertex 
functions are calculated using a dimensional regularized scheme with
minimal subtraction of poles~\cite{Am}. 
The vertex function $\Gamma^{(2)}$ yields the wave
function renormalization $Z_{\Phi}=1+O(u^{2})$. The determination of
$Z_{u}$ is obtained by the $1/\epsilon$-part of $\Gamma^{(4)}$ and is 
\begin{equation}
\label{b10}
Z_{u}=1+\frac{3u}{2\epsilon}+O(u^{2})\quad ,
\end{equation}
from which follows the $\beta$-function the zero of which yields the
infra-red stable fixed point $u^{*}=2\epsilon/3+O(\epsilon^{2})$. An evaluation
of $\Gamma^{(2,1,0)}$ yields
\begin{equation}
\label{b11}
Z_{\Phi^{2}}=1+\frac{u}{2\epsilon}+O(u^{2})\quad ,
\end{equation}
whereas $\Gamma^{(2,0,1)}$ yields
\begin{equation}
\label{b12}
Z_{B}=1+\frac{u}{3\epsilon}+O(u^{2})\quad .
\end{equation}
From the above equations we now obtain the exponents due to the 
relations~\cite{Am&G,Am}
\begin{eqnarray}
\label{b13}
\eta = \gamma_{\Phi}(u^{*},\epsilon),\;\;
\nu^{-1} -2 = \gamma_{\Phi^{2}}(u^{*},\epsilon),\; \;
\nu_{B}^{-1} -2 = \gamma_{B}(u^{*},\epsilon)\quad ,
\end{eqnarray}
which yield the exponents given in equations~(\ref{b3})
and~(\ref{b4}). 
Our results are to order
$\epsilon$, with corrections of the order of $\epsilon^{2}$.

To see the scaling form of $<R^{2}>$ we need to solve the RG equation for
the Fourier transformed chain partition function 
$\tilde Z_{R}^{(2)}(q)=T^{-1}_{{\cal L}}(\tilde G^{(2)}(q))$ which is given by
\begin{equation}
\label{b14}
\left[
\mu\partial_{\mu}+\beta\partial_{u}
+\gamma_{\Phi^{2}}\tau\partial_{\tau}-\gamma_{B}L\partial_{L}
+\gamma_{\Phi}+\gamma_{B}\right]
\tilde Z_{R}^{(2)}(q,\tau,L,u,\mu) =0\quad ,
\end{equation}
The solution is 
obtained by the method of characteristics~\cite{Am} and gives the scaling form
of $<R^{2}>$
to be~\cite{SV}
\begin{eqnarray}
\label{b19}
F(\tau ,{\cal L})&=& {\cal L}^{2\nu_{B}} \tilde 
F(\tau {\cal L}^{\nu_{B}/\nu},u^{*})\nonumber\\
          &=& |\tau|^{-2\nu} \tilde 
F_{\pm}({\cal L} |\tau |^{\nu/\nu_{B}},u^{*})
\end{eqnarray}
We see that in the above equations the cross over exponent $\phi$
is appeared. Here $\pm$ denotes $T>T_{c}$ and $T<T_{c}$, respectively.

Now we calculate the explicit form of the scaling function of the 
end to end distance $<R^{2}>$  in case 
of $\tau \to 0$ ($T\ge T_{c}$) and
${\cal L}\to \infty$ 
so that $\tau {\cal L}$ stays finite. We need the general
matrix renormalization scheme, because no soft expansion wrt the
monomer chemical potential
is possible. We have to use instead of equation~(\ref{b6}) for the
vertex renormalization 
\begin{eqnarray}
\label{b20}
y_{0} &=& Z_{21}\tau + Z_{22}y \nonumber\\
\tau_{0} &=& Z_{11}\tau + Z_{12}y \quad .
\end{eqnarray}
These $Z$-factors are simply given by the renormalization of the 
corresponding vertex functions 
with zero momentum for the polymer and the fluid
$\tilde \Gamma_{22}^{(2)}(k=0)$ and
$\tilde \Gamma _{11}^{(2)}(k=0)$ which fix the values of $Z_{21}$ and $Z_{22}$
and $Z_{11}$ and $Z_{12}$.
They   
are given to order 1-Loop in the figures 1 and 2. 
\begin{figure}
\begin{center}
\begin{minipage}{11cm}
{{\epsfig{file=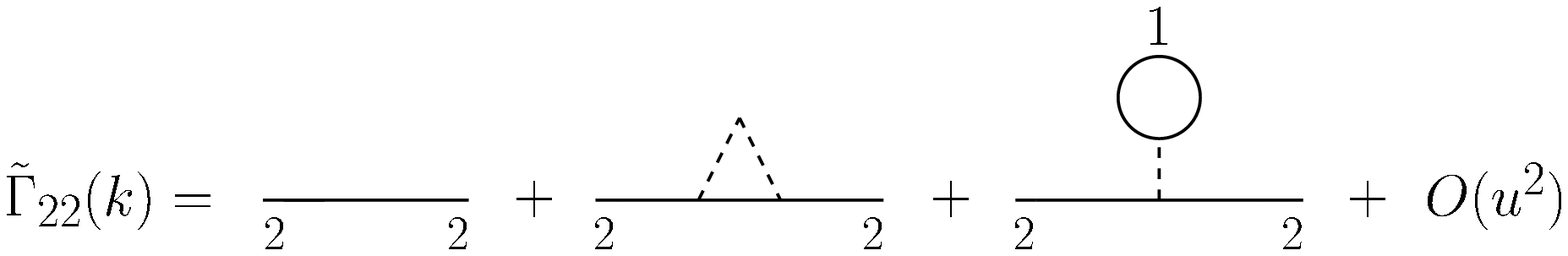,width=7cm}}}
\caption{{\small {The diagrams for the polymer $\Gamma^{(2)}_{22}$. 
A line with the number 2 corresponds to a free polymer propagator
$(k^{2}+y)^{-1}$, a line with 1 to a 
fluid propagator $(k^{2}+\tau)^{-1}$ and doted line
to a interaction vertex $u/4!$. 
Note that we do not include external legs and evaluate the
graphs at $k=0$ for the vertex
function. The chain propagator $G^{(2)}$ is also calculated using
these graphs 
including external legs and with $k$ finite.}}}
\end{minipage}
\end{center}
\end{figure}
\begin{figure}
\begin{center}
\begin{minipage}{11cm}
\centerline{{\epsfig{file=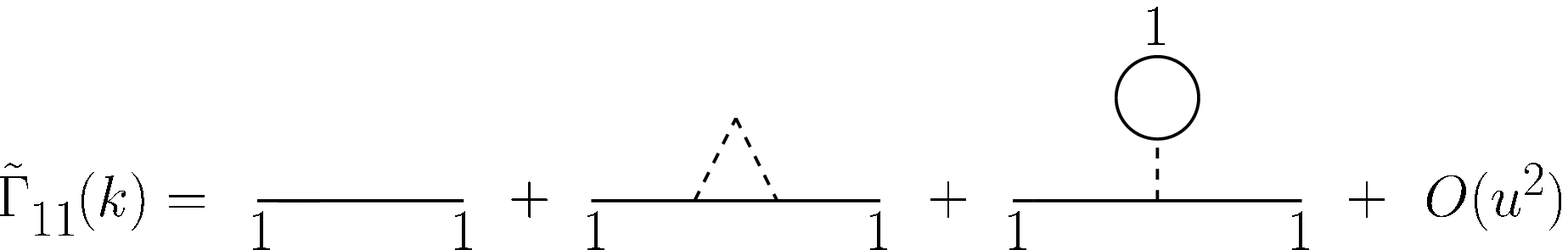,width=7cm}}}
\caption{{\small {The diagrams for the fluid $\Gamma^{(2)}_{11}$}}}
\end{minipage}
\end{center}
\end{figure}
The evaluation of their Feynman 
diagrams yield $Z_{21}=\frac{u}{6\epsilon}+O(u^{2})$,
$Z_{22}=1+\frac{u}{3\epsilon}+O(u^{2})$, 
$Z_{11}=1+\frac{u}{2\epsilon}+O(u^{2})$ and
$Z_{12}=0$, in the polymer limit $M\to 1$. 

The corresponding RG equation for $\tilde \Gamma_{{22R}}^{(2)}$ is then
\begin{equation}
\label{b211}
\left[\mu\partial_{\mu}+\beta\partial_{u}+\kappa_{\tau}\tau\partial_{\tau}+
(\kappa_{21}\tau+\kappa_{22}y)\partial_{y}
-\gamma_{\Phi}\right]\tilde\Gamma^{(2)}_{22R}=0\quad ,
\end{equation}
and follows from a variation wrt the external length $\mu^{-1}$. The new
coefficients are $\kappa_{\tau}=-\beta \partial_{u}\ln{Z_{11}}$,
$\kappa_{22}=-\beta\partial_{u}\ln{Z_{22}}$ and 
$\kappa_{21}=-1/Z_{22}(\beta\partial_{u}
Z_{21}-Z_{21}\kappa_{\tau})$.

After an inverse Laplace transformation wrt $y$ of the RG equation for the
propagator $\tilde G^{{(2)}}=1/\tilde\Gamma^{(2)}$ we get for the
2-point chain partition function
\begin{equation}
\label{b21}
\left[\mu\partial_{\mu}+\beta\partial_{u}+\tau\kappa_{\tau}\partial_{\tau}
-\kappa_{22} L\partial_{L}
+\gamma_{\Phi}-\kappa_{22}-\kappa_{21}\tau L\right]
\tilde Z^{(2)}_{22R}=0\quad .
\end{equation}
With the use of characteristics as mentioned 
after~(\ref{b14}) we find the solution
of~(\ref{b21}) and it turns out
the same scaling form for $< R^{2} >$ as in the soft 
expansion~(\ref{b19}) because both renormalization schemes are equivalent.
But the critical
exponents are now defined as $1/\nu_{B}=2-\kappa_{22}(u^{*})$,
$1/\nu=2-\kappa_{\tau}(u^{*})$ and $\eta =\gamma_{\Phi}(u^{*})$. 
Because the solution of the general vertex matrix renormalization has to be
consistent with the diagonal (soft expansion) scheme of~(\ref{b14}), it follows
that $Z_{22}\equiv Z_{B}$ and $Z_{11}\equiv Z_{\Phi^{2}}$ to all orders in $u$.
But we need the matrix scheme for the analysis of the
explicit loop-expansion: By using the flow equations from the method of
characteristics
 similar as in the calculation of the equation of state for magnetic
systems~\cite{Am} we get from the expansion of $\tilde \Gamma^{{(2)}}_{22}$
finally the explicit result valid to order $\epsilon$
\begin{eqnarray}
\label{b23}
<R^{2}> &\sim&{\cal L}^{2\nu_{B}} \quad .
\end{eqnarray}
This 
scaling form is consistent with the general result~(\ref{b19}).

Now we express our renormalized quantities by the original bare ones and
find that
\begin{equation}
\label{b24}
<R^{2}>= D(u) l^{2}N^{2\nu_{B}}
\quad ,
\end{equation}
with a non universal constant $D(u)$~\cite{SV}. 
There is in 1-loop order no explicit
temperature dependence of the demixing binary solvent but we have the
implicit condition that $\tau_{0}Nl^{2}$ is finite in the critical limit.

To summarize: we found that in the critical region the demixing
tendency of the fluid yields, if the
 number of monomers $N\to \infty$ and the critical temperature deviation of
the fluid $\tau_{0}\to 0$ such that $lN\tau_{0}$ stays finite, to a
smaller chain size than in each of the good solvent components separately. 
This is 
because the new exponent $\nu_{B}$ is less than the self avoiding walk
exponent $\nu_{0}=(1+\epsilon/8+O(\epsilon^{2}))/2$. 
To get an explicit demixing temperature dependence 
in the end-to-end distance of the polymer chain a calculation
to a higher Loop order is necessary. Now only the exponent $\nu_{B}$ 
can be found in 2-Loop order in reference~\cite{Am&G}.

The binary solvent with preferential affinity to the monomers is not 
tractable by the renormalization group. Due to the induced preferential
adsorption the chain likes to be surrounded by the better solvent. This
effective long range attraction of the range of the fluid correlation length
yields in a specific temperature range 
to a partly collapsed chain~\cite{1-4,Br&G,Ma,Vilgis}. In the
RG calculation this manifests in a dominant cubic term in the Hamiltonian which
implies an imaginary solution of the $\beta$-function~\cite{SV}.

A related problem is a screened polyelectrolyte chain in a demixing
binary solvent. The effective attractive interaction induced by the
preferential affinity leads then to an effective charge of the 
monomers~\cite{SV}. 

Real systems have preferential affinity. Therefore  its interesting to
investigate if at least right at the consolute point of the fluid the
preferential adsorption term has still significant influence on the 
(infinite) chain size, because the chain is then again swollen. This could
be done by computer-experiments.

{\em Note added}: After acceptance of the present paper an
  experimental study appeared \cite{exc} which agrees in several points with
our results.


We acknowledge helpful discussions with H. K. Janssen, T. M. Liverpool
and \mbox{J. J. Magda}.
We particularly thank H. K. Janssen and T. M. Liverpool for a critical
reading of the manuscript.


\begin{figure}
\epsfxsize 12cm
\epsffile{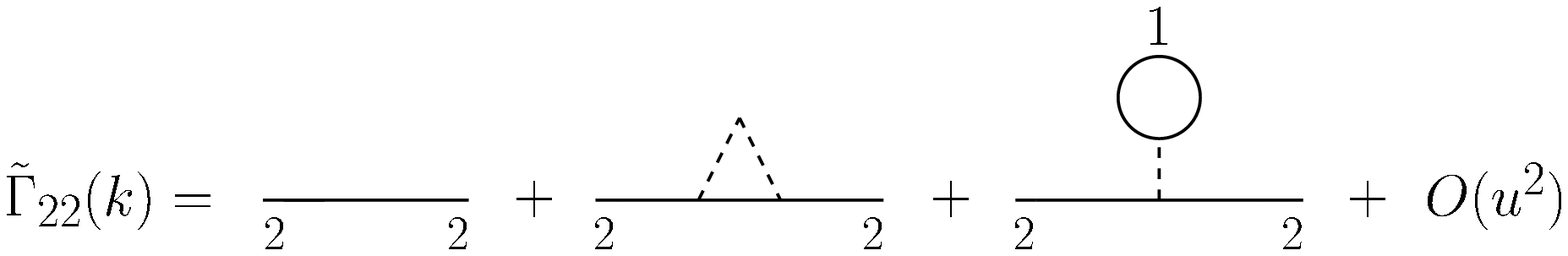}
\caption{The diagrams for the polymer $\Gamma^{(2)}_{22}$. 
A line with the number 2 corresponds to a free polymer propagator
$(k^{2}+y)^{-1}$, a line with 1 to a 
fluid propagator $(k^{2}+\tau)^{-1}$ and doted line
to a interaction vertex $u/4!$. 
Note that we do not include external legs and evaluate the
graphs at $k=0$ for the vertex
function. The chain propagator $G^{(2)}$ is also calculated using
these graphs 
including external legs and with $k$ finite.}
\label{fig1}
\end{figure}
\begin{figure}
\epsfxsize 12cm
\epsffile{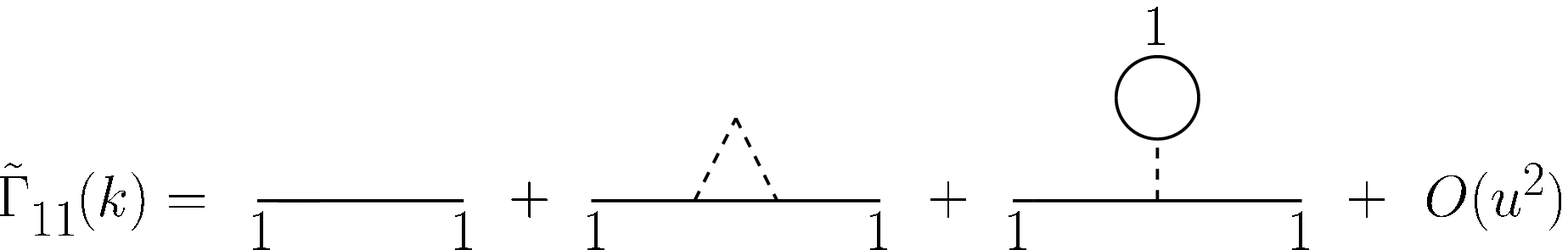}
\caption{The diagrams for the fluid $\Gamma^{(2)}_{11}$.}
\label{fig2}
\end{figure}

\vskip-12pt
\begin{thebibliography}{99}
\bibitem{1-4}
R. C. Schulz and P. J. Flory
\newblock {\em 1.\ .Polym.\ Sci.}, {\bf 15}\ 231, (1955);
P. G. de Gennes \newblock {\em J. \ Phys.\ France}, {\bf 37}\ 59, (1976);
A. Dondos and H. Benoit
\newblock {\em Makromol.\ Chem.}, {\bf 133}\ 119, (1970).
\bibitem{Br&G}
F. Brochard and P. G. de Gennes
\newblock {\em Ferroelectrics}, {\bf 30}\ 33, (1980).
\bibitem{Ma}
J. J. Magda, G. H. Fredrickson, R. G. Larson and E. Helfand
\newblock {\em Makromolecules}, {\bf 23}\ 726, (1988).
\bibitem{Vilgis}
T. A. Vilgis, S. Sans and G. Jannink
\newblock {\em J.\ Phys.\ France}, {\bf 3}\ 1779, (1993)
\bibitem{Doi}
M. Doi and S. E. Edwards
\newblock {\em The Theory of Polymer Dynamics}, 
(Oxford University Press, Oxford, 1976)
\bibitem{Am&G}
D. J. Amit and Y. Y. Goldschmidt
\newblock {\em Annals of Physics}, {\bf 114}\ 356, (1978)
\bibitem{Am}
D. Amit, \newblock {\em Field Theory,  the
 Renormalization Group and Critical Phenomena}, (World Scientific, Singapore,
1984); J. Zinn-Justin,
\newblock{\em Quantum Field Theory and Critical Phenomena},  (Oxford
University Press, Oxford, 1996); E. Eisenriegler, 
\newblock {\em Polymers near Surfaces}, (World Scientific, Singapore, 1993)
\bibitem{LS}
T. B. Liverpool and M. Stapper \newblock {\em Europhys. Lett.} 
{\bf 40}, 485 (1997)
\bibitem{SV}
M. Stapper and T. A. Vilgis \newblock {\em in preparation} (1997)
\bibitem{exc}
K. To and H.J. Choi \newblock {\em Phys. Rev. Lett.}, {\bf 80}, 536, (1998)
\end{thebibliography}
\end{document}